# Mining Social Media to Inform Peatland Fire and Haze Disaster Management

**Mark Kibanov** · **Gerd Stumme** ·
**Imaduddin Amin** · **Jong Gun Lee**

**Abstract** Peatland fires and haze events are disasters with national, regional, and international implications. The phenomena lead to direct damage to local assets, as well as broader economic and environmental losses. Satellite imagery is still the main and often the only available source of information for disaster management. In this article, we test the potential of social media to assist disaster management. To this end, we compare insights from two datasets: fire hotspots detected via NASA satellite imagery and almost all GPS-stamped tweets from Sumatra Island, Indonesia, posted during 2014. Sumatra Island is chosen as it regularly experiences a significant number of haze events, which affect citizens in Indonesia as well as in nearby countries including Malaysia and Singapore. We analyse temporal correlations between the datasets and their geo-spatial interdependence. Furthermore, we show how Twitter data reveals changes in users' behavior during severe haze events. Overall, we demonstrate that social media are a valuable source of *complementary and supplementary* information for haze disaster management. Based on our methodology and findings, an analytics tool to improve peatland fire and haze disaster management by the Indonesian authorities is under development.

M. Kibanov, G. Stumme
Knowledge & Data Engineering Group
ITeG Research Center
University of Kassel, Germany
E-mail: kibanov, stumme@cs.uni-kassel.de

I. Amin, J.G. Lee
Pulse Lab Jakarta
UN Global Pulse
United Nations
E-mail: imaduddin.amin, jonggun.lee@un.or.id

## 1 Introduction

Peatland fires and their associated haze events are slow onset but medium impact disasters. At times the fires occur when several environmental conditions meet, but most fires are man-made disasters resulting from – predominantly illegal – agricultural practices, e. g., conversion often forests and peatland into palm oil plantations through slash-and-burn techniques [17]. The impact of these activities span environmental (e. g., accelerated deforestation by up to 62% [23]) and economic losses, but the associated haze events also seriously affect local residents' health. When such fires, in particular peat fires, produce haze, wide areas can be affected within a relatively short period of time. If the haze severity level is low, the Government of Indonesia advises its citizens to reduce their outdoor activities. If the level is high, residents are asked to evacuate the affected area, because haze generated by peatland fires may lead to various health issues [1],[14]. Dense haze, including a case in Indonesia on October 20, 2015 as shown in the image below reduces the visibility to less than 100 meters (and in rare instances to less than 20-30 meters) and leads the closure of airports and schools. Haze is not only a national issue for Indonesia, but also an international issue, as it affects Singapore and Malaysia.

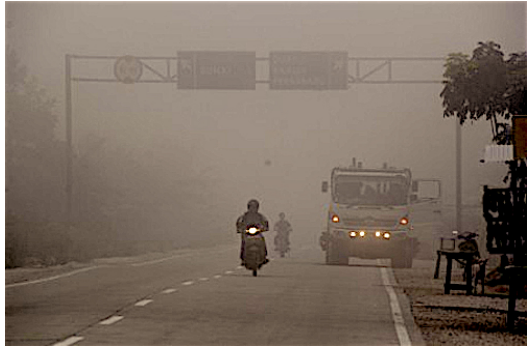

*"Riau extends haze emergency status"* (**Source - Jakarta Post**[1])

In order to respond to a fire or haze event, a disaster management authority needs fire hotspot information along with (static) baseline information, including an estimate of the affected population and data on the available facilities. Fire hotspots (henceforth referred to as 'hotspot' or 'hotspots') are identified relatively efficiently from satellite imagery using a classification algorithm. Due to the limited resources available to public authorities, more efficient and effective approaches to disaster management are welcome, including through the generation of new information.

In this article, we investigate the opportunities that social media offer for improved management of peatland fire and haze disasters. Specifically we analyse Twitter as the primary data source and use hotspot and air quality data to further interpret and verify the results of our analysis. Our objective is not the evaluation of informa-tion offered by other channels, such as hotspots identified from satellite imagery, but the provision of 'complementary' situational information. We concentrate our analysis on peatland fires on Sumatra Island as the haze situation in this region is the most critical in terms of geographic coverage and severity – as explained in Figure 1, and, in doing so, analyse (almost) all geotagged tweets from Sumatra Island posted during 2014.

---

[1] http://www.thejakartapost.com/news/2015/10/20/riau-extends-haze-emergency-status.html (accessed on Jan 12, 2016)

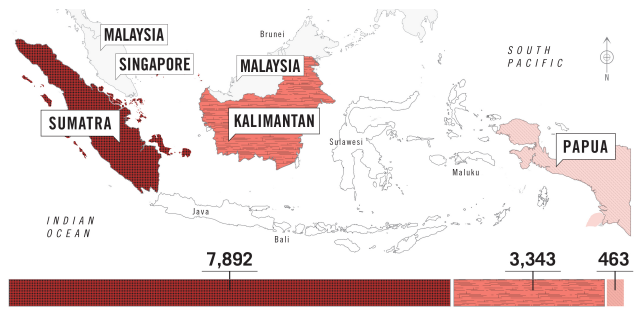

**Fig. 1: The three islands in Indonesia (Sumatra, Kalimantan, and Papua) with the largest numbers of hotspots in 2014**

We aim to explore two research themes concerning the utility of social media for haze disaster management.

1. How do peat fires/haze events and social media conversations about haze-related topics relate to each other? If such a (temporal and geo-spatial) relationship exists, how can it be described and quantified?
2. How do social media capture situational information on affected populations, such as mobility patterns, which can be useful in managing disaster response?

The contribution of our work with regard to haze disaster response on Sumatra Island can be summarized as follows:

1. We find strong temporal and geo-spatial correlations between hotspots and haze-related conversations on Twitter.
2. We investigate the mobility patterns of haze-affected populations as captured in social media data and confirm that they reflect movement patterns during haze situations in the real-world.

These insights can significantly contribute to the prioritisation of haze disaster response activities, conducted by national and local public authorities in Indonesia, by raising real-time awareness of the social impact of the phenomenon, as opposed to current practice which uses hotspot and population data in isolation.

The rest of this article is structured as follows. In Section 2 we explain related work. Section 3 discusses the data we use for our analysis, covering both insights and limitations. Section 4 presents how different online conversations reflect real-world haze disasters. Section 5 shows what kind of information on population mobility can be developed from social media with respect to haze disasters. We close this article with a discussion on research issues in Section 6 and a conclusion in Section 7.

## 2 Related Work

We focus on two groups of related work – studies of how social media can be used to assist disaster management and research about data mining studies on forest fires.

**Social Media and Disasters**: A vast body of research exists on social media and its potential reflection of real-world phenomena. We consider the literature that demonstrates how social media relate to different disasters.

Carley *et al.* [12] discuss the usage of Twitter for disaster management in Indonesia. This work makes two important contributions relevant for the current article. First, the viability of using Twitter data for disaster management is demonstrated; second, the paper provides a baseline for Twitter use to support disaster management.

The potential application of social media to disaster and crisis management is attracting the research community's attention either as a tool or as a source of data, *e.g.,* for the creation of crisis maps. Oz and Bisgin [30] made a social-media based research about attribution of responsibility (e.g., how users assign political responsibility) regarding Flint water crisis. Gao *et al.* [15] consider social-media-based crowdsourced maps with data from external sources as a powerful tool in humanitarian assistance and disaster relief. Goolsby [16] describes how social media can be used as crisis management platform to create crisis maps for different agencies. Middleton *et al.* [28] propose a social media crisis-mapping platform, where real-time crisis maps are generated based on statistical analysis of tweet streams matched to areas at risk. Cresci *et al.* [13] propose a crisis mapping system that overcomes some limitations of other systems: they introduce an SVM-based damage detection approach, and describe a new geoparsing technique to perform a better geolocation of social media messages.

Imran *et al.* [19] provide an overview of existing and proposed methods and systems to retrieve information about emergencies from social media, such as Crisis-Tracker [33] and TweetTracker [26]. Furthermore, Imran *et al.* suggest a platform to collect human annotations in order to maintain automatic supervised classifiers for social media messages [20] and describe automatic methods for extracting brief, self-contained information items from social media, which are relevant to disaster response [21]. Abel *et al.* [7] propose the 'Twitcident', a system for filtering, searching and analyzing information about real-world incidents or crises. A number of systems are implemented and deployed for special kinds of disasters: Avvenuti *et al.* [9][10] describe the design, implementation and deployment of social media based system for detection and damage assessment of earthquakes in Italy. The system is able to detect outbraking seismic events with low occurrences of false positives. Cameron *et al.* [11] describe ESA-AWTM – a system deployed for trial by the Australian Government. This tool formalises the usage of Twitter by the Crisis Coordination Center, replacing an ad-hoc fashion monitoring of social media.

Following studies focus rather on algorithmical aspects of social media for disaster management, rather than on system architecture or implementation. Rogstadius [32] discusses different aspects of possibilities for enhanced disaster situational awareness using social media. Sakaki *et al.* [34] present how social media can be utilised as an early warning system with regard to earthquake events. Krstajic *et* *al.* [24] show not only how natural disasters and man-made catastrophes can be detected using Twitter, but also how semantic aspects of such events can be represented. Schulz *et al.* [36] explore the possibilities of detection of small scale incidents using microblogs. Mandel *et al.* [27] made a demographic analysis of online sentiment during a hurricane, particularly focusing on the level of concern disaggregated by gender. Morstatter *et al.* [29] address the problem of finding (non-geotagged) tweets that originate from a crisis region. Zhang and Vucetic [38] propose an improved method to identify disaster-related tweets using a semi-supervised approach with an unlabeled corpus of tweets.

Disaster damage assessment is another important topic in context of disaster management: Kryvasheyeu *et al.* [25] examine the online activity of different areas before, during and after Hurricane Sandy. They demonstrate that per-capita Twitter activity strongly correlates with the per-capita economic damage inflicted by the hurricane.

**Forest Fires/Haze and Data Mining**: The following works contribute to the forest fires issue, applying data mining methods to different aspects of this emergency. We do not address in detail the topic of forest fire detection using satellite imagery as this topic is not in focus of current work, but we note that usage of satellite imagery for forest fires detection is pretty well studied; *e.g.*, Jaiswal *et al.* [22] describe how forest fire risk zones can be mapped using satellite imagery and geographic information systems.

A number of studies present novel data mining methods for haze-related issues. Sakr *et al.* [35] presents a model to predict forest fires risks using data on previous weather conditions with the best results having been achieved using support vector machines and a Gaussian kernel function. Iliadis [18] introduce a decision support system for long-term forest fire risk estimation, based on fuzzy algebra. The system was applied in Greece but, according to authors, can be used on a global basis. Sitanggang and Ismail [37] suggest a classification model for hotspot occurrences using a decision tree model C4.5 algorithm, which achieves an accuracy of 63%. The forest fire data from the Rokan Hilir district on Sumatra Island, Indonesia was used in that research.

Prasetyo *et al.* [31] used Twitter and Foursquare data to analyze public perceptions of haze in Singapore. In particular, their analysis showed that (1) social media users focus significantly on the haze problem and (2) the problem has a substantial emotional and behavioral impact. The UN presented the first feasibility study [5] for supporting peat fire and haze disaster management using social media.

**Overall**, there are a lot of publications describing different aspects of social media as an assistance tool for disaster management at different scales. Some researchers applied data analysis and mining techniques to forest and peatland fires data.

In this article, we concentrate on peatland fires in Indonesia. This disaster occurs periodically and has a huge impact on environment and lives of millions of people. In contrast to previous works, we consider the dataset of (almost) all geotagged tweets and all hotspots during one year on Sumatra Island. We created four rich taxonomies for identification of different topics related to haze and peatland fires. Furthermore, one of the main focuses of our work is to estimate whether social media can be used for estimation of users' behaviour w.r.t. Twitter and mobility. To the best of the authors' knowledge, those aspects were not covered in previous works.

## 3 Data Set

We use two main datasets: twitter data from Sumatra Island and hotspots on Sumatra Island detected by National Aeronautics and Space Administration (NASA) satellites. In this section, we describe the data, its basic characteristics and its limitations.

### 3.1 Fire Hotspots and Twitter Data

**Fire Hotspots in Sumatra in 2014**: We use hotspot information that is first identified by the NASA from imagery, captured by Terra and Aqua satellites[4], and further refined and augmented by GLOBAL FOREST WATCH[2] [3]. Based on consultations with and recommendations by a domain expert from the UN Office for REDD+ Coordination in Indonesia (UNORCID)[3], we filter hotspots classified as 'peatland hotspot' and 'high confidence-level' between two confidence levels (*i.e.*, high and low) and use 7,892 hotspots for our analyses after the filtering process, among the 38,723 hotspots in Sumatra in 2014 discovered by GLOBAL FOREST WATCH, cf. Table 1.

**Table 1: Size of hotspot and Twitter datasets**

| Area | Fire Hotspot | | | Twitter |
|---|---|---|---|---|
| | All | Peatfire Only | Peatfire & High Conf. | Tweet |
| Sumatra Island | 38,723 | 27,060 | 7,892 | 29,528,786 |
| Riau Province | 21,563 | 19,191 | 5,463 | 3,509,849 |

**(Almost) All GPS-stamped tweets posted in Sumatra during 2014**: Indonesia has a large number of Twitter users, which stands at approximately 12 million in 2014 [2]. Originally we attempt to collect the entire GPS-stamped tweets posted within Sumatra Island between January 1 and December 31, 2014 but for unexpected technical reasons, we failed to collect tweets during two time frames:

- about half of all tweets posted during January 2014
- no tweets collected between April 16 and 30 2014

Overall, we analyse more than 29 million tweets with exact locational information, which includes longitude and latitude. The tweets are posted by 575,295 users, equivalent to 1% of the population of Sumatra Island.

---

[2] http://www.globalforestwatch.org

[3] UN REDD Programme is United Nations Collaboration for Reducing Emissions from Deforestation and Forest Degradation in Developing Countries

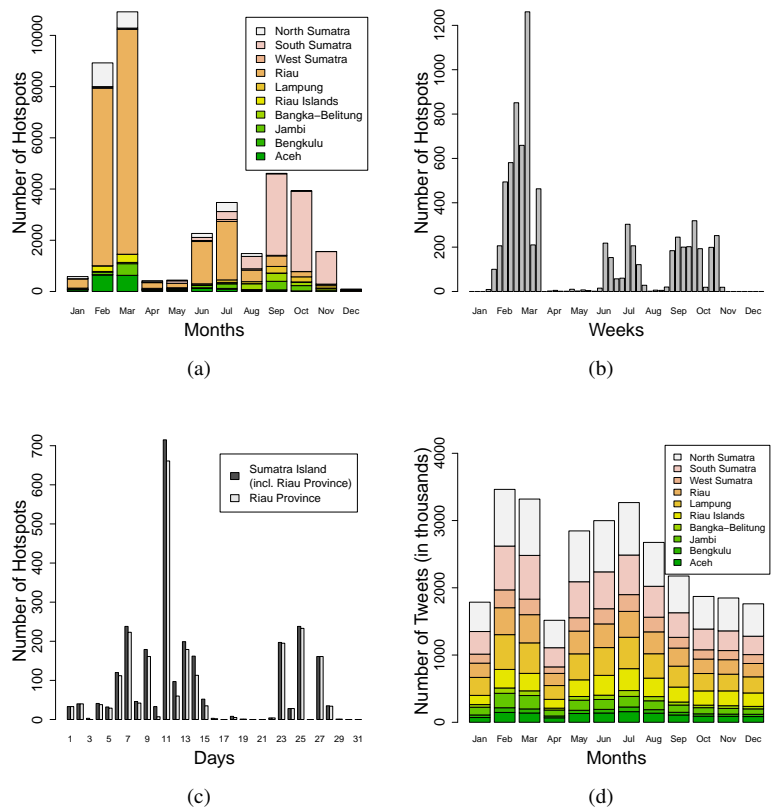

**Fig. 2: Basic characteristics of hotspot and tweet datasets - (a) Monthly hotspots, by province, on Sumatra in 2014, (b) weekly hotspots on Sumatra in 2014, (c) daily hotspots on Sumatra island and in Riau province in March 2014, and (d) monthly tweets by province of Sumatra in 2014 (a) and (d) are best viewed in color)**

### 3.2 Basic characteristics

**Background**: Sumatra Island is an area heavily affected by peatland fire and haze events, as shown in Figure 1. The central and southern parts of the island experience more haze than other areas and in particular, Riau province is recognised as the most haze prone area, not only based on the total yearly hotspots, as shown in Figure 2(a), but also based on the number of haze situations reported in the media.

**Hotspots**: Among other interesting observations, about 5,500 high confidence peat fire hotspots are detected in Riau Province during 2014, while the number of high confidence peat fire hotspots across Sumatra stands at 7,892 in 2014, which means almost 70% of high confidence peat fires take place on less than 19% of the island area, namely Riau province. This leads us to a further investigation, namely of the effects of haze on the behaviors of people who live in Riau province, the results of which will be discussed in Section 5. It is worth mentioning that there are two main fire periods on Sumatra Island in 2014: between February and March, and between June and November as shown in Figures 2(a) and 2(b). In particular, Figure 2(b), the weekly hotspot dynamics shows a similar pattern to the monthly pattern provided in Figure 2(a). However, we can find it is similar to monthly dynamics, but there are some less severely affected weeks in June, July and October and strong irregularity in different scales. Moreover, when we have a close look at hotspots dynamic in March, the worst month in 2014 (c.f. Figure 2(c), we find that the hotspots almost disappeared after evacuation announcement but then emerged again after one week.

**Tweets**: Comparing Figure 2(d) and Figure 2(a) highlight that the dynamics pattern from Twitter use in general is different from the patterns of hotspots. The usage of Twitter on Sumatra Island decreases towards the end of 2014, and the distribution of tweets among the different provinces remains stable during the year.

### 3.3 Data limitations

This section briefly elaborates four limitations of social media data, the first two concern inherent limitations of social media and the other two concern the limitations of our data in particular. First, the digital divide may introduce a limitation, in that social media data connected to urban areas are considerably denser compared with rural areas and in that the cost of smart phones limits the participation of less affluent cohorts of society. Secondly, there may be some bias connected to the characteristics of users of social media in that we largely 'hear' the opinions of younger or more extrovert users. Thirdly, the number of unique users from our dataset (about 500,000 users) equates to approximately 1% of the entire population of Sumatra island; it thus may also limit our ability to abstract or generalise based on our findings. Fourthly, due to technical issues, our data are incomplete, specifically in January and April 2014.

It is, however, worth noting that a key objective of this work is to test the potential of social media as a complementary data source which can be used to inform humanitarian efforts. The aim is not to produce statistics solely from social media but to sense in near real-time signals on the behaviors of affected populations, which could better inform decision-making and improve the targeting of humanitarian response. We expect that the missing data will not affect the validity of our analysis because the missing periods are predominantly outside of the observed haze seasons.

## 4 Temporal and spatial analysis

In this section, we investigate the temporal and spatial characteristics of tweets related to haze and for that, prepare four different datasets by developing four taxonomies related to haze to identify four different haze-related topics, *i.e.,* `haze-general`, `haze-hashtag`, `haze-impact`, and `haze-health`. We extend a simple taxonomy used for a feasibility study [5] in this work in order to capture broader contexts related to haze crises. Overall, we establish 126 filtering rules, combinations of Indonesian words and boolean operators as presented in Table 2, which identify 64,383 tweets posted by 33,127 users out of 29,528,786 tweets across four topics, while allowing for overlapping topics. For ease of understanding, even though we use both `||` and `OR` in different places, both operators have one meaning, namely `OR`. For instance, the following rule will collect a tweet if its content contains both `A` and `C` or both `B` and `C`, [Rule] - " **(** `A ||` `B` **)** `&& C` ".

**Table 2: Four types of conversations related to peat fires and haze in 2014 on Sumatra Island and filtering rules for the identification of corresponding tweets - `haze-general` (43), `haze-hashtags` (5), `haze-health` (39), and `haze-impact` (39)**

| | |
|---|---|
| **`haze-general`** | **Conversations about forest and peat fires and haze, detected primarily by the keywords —** 9,707 tweets (e.g., *"When the haze problem will be solved?"*): <br> `( (bencana||badai||polusi||parah||tebal||kabut) && (asap||kabut) )` <br> `OR ( (awas||berbahaya||darurat||pekat||lebat) && (asap||kabut) )` <br> `OR ( kabut asap||titik api||sumber api||titik panas||polusi udara||haze )` <br> `OR ( forest fire ) OR ( (kerusakan||pembalakan||pembukaan||kebakaran ||penggundulan||penebangan) && (hutan||ladang||lahan||gambut) )` |
| **`haze-hashtag`** | **Conversations which contain one of identified hashtags — 3,024 tweets** <br> (e.g., *"Let's participate in #melawanasap movement."*): <br> `( #saveriau||#prayforriau||#melawanasap||#prayforasap||#hentikanasap )` |
| **`haze-impact`** | **Conversations about happenings in a negative way due to haze, such as flight delay or school closing — 6,994 tweets** (e.g., *"Day #3 off because of Haze."*): <br> `( (tutup||batal||dibatalkan||tertunda||delay||cancel||ditutup) && (penerbangan||bandara) ) OR ( jarak pandang )` <br> `OR ( (sekolah||kampus||kuliah) && (tutup||ditutup||libur||diliburkan) )` <br> `OR ( (ekonomi||dampak||akibat||merugi||lumpuh||resiko) && (asap||kabut) )` |
| **`haze-health`** | **Conversations with keywords indicating haze-related or derivable diseases —** 46,241 tweets (e.g., *"Welcome to Pekanbaru; do not forget to wear mask!"*): <br> `( (infeksi||sesak) && (pernapasan||napas||pernafasan) )` <br> `OR ( (iritasi||radang) && (mata||kulit||enggorokan||hidung||paru) )` <br> `OR ( batuk||pusing||mual||ispa||masker||asma||asthma||paru-paru )` <br> `OR ( (asap||kabut) && (kesehatan||sehat||pernafasan||hamil||anak ||orang tua) ) OR ( mata && (pedih||perih||sakit) )` |

### 4.1 Temporal Dynamics of Tweet Topics and Hotspots

Users, inherently, engage in more intensive discussions on haze not only offline but also online, when the haze situation deteriorates, which is quantitatively measured by the number of hotspots.

In order to visualize any correlation between two quantities of haze-relevant conversations and haze situations, we present the weekly numbers of tweets from four datasets as well as the number of hotspots on Sumatra island in Figure 3. Although our data are incomplete in January and April due to technical issues, as explained in the previous section, the figure shows that Twitter users respond to haze events on the ground by discussing on social media the four topics of interest.

**Table 3: Values of correlation coefficients between the weekly number of hotspots and the number of tweets on Sumatra Island and in Riau Province**

| Area | haze-general | haze-hashtag | haze-impact | haze-health |
|---|---|---|---|---|
| Sumatra | 0.89 | 0.71 | 0.83 | 0.79 |
| Riau | 0.89 | 0.72 | 0.85 | 0.91 |

To easily quantify such correlations, we calculate Pearson's correlation between tweet dynamics and hotspot dynamics in Sumatra and it reveals strong correlation coefficient values as shown in Table 3. Among the four conversation topics, `haze-general` has the strongest coefficient. When we conduct the same calculation with tweets and hotspots in Riau, it also shows strong correlation coefficients but in this instance `haze-health` is strongest, which is understandable because Riau is one of the most haze-affected provinces in Sumatra island.

The high values of correlation coefficients confirm that four identified conversations are relevant for haze problems. This is a confirmation of relevancy of chosen taxonomies, Table 2.

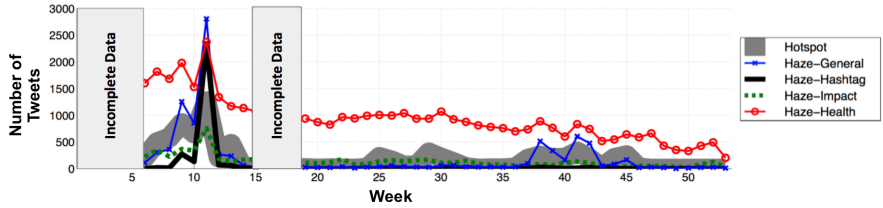

**Fig. 3: Correlation between the weekly number of hotspots and the weekly conversations about different topics on Sumatra Island**

In Figure 4, per topic we plot weekly tweet volume and hotspot volume in $x$ and $y$ axes, respectively, while all $y$ axes are presented in a logarithm scale. As already shown in Figure 3 and Table 3, we find positive correlations from four subfigures but confirm an interesting phenomenon that people discuss the studied topics except `haze-hashtag` during haze-free periods. The conversations classified by `haze-hashtag` happened only during haze periods. This behaviour is explained by the observation that specific hashtags are commonly used in connection to a specific event.

### 4.2 Spatial Characteristics of Different Topical Conversations

This section discusses the spatial relationships between hotspots and tweets. For instance, if one can identify hotspots (not areas) from social media, it would be useful for the prioritisation of disaster response and humanitarian action during haze disasters. It is still challenging to understand to which hotspot a user refers in a tweet for many reasons. We investigate spatial characteristics using a simple mapping process based on the distance between the position of a tweet and a hotspot. This still allows for interesting insights on spatial characteristics.

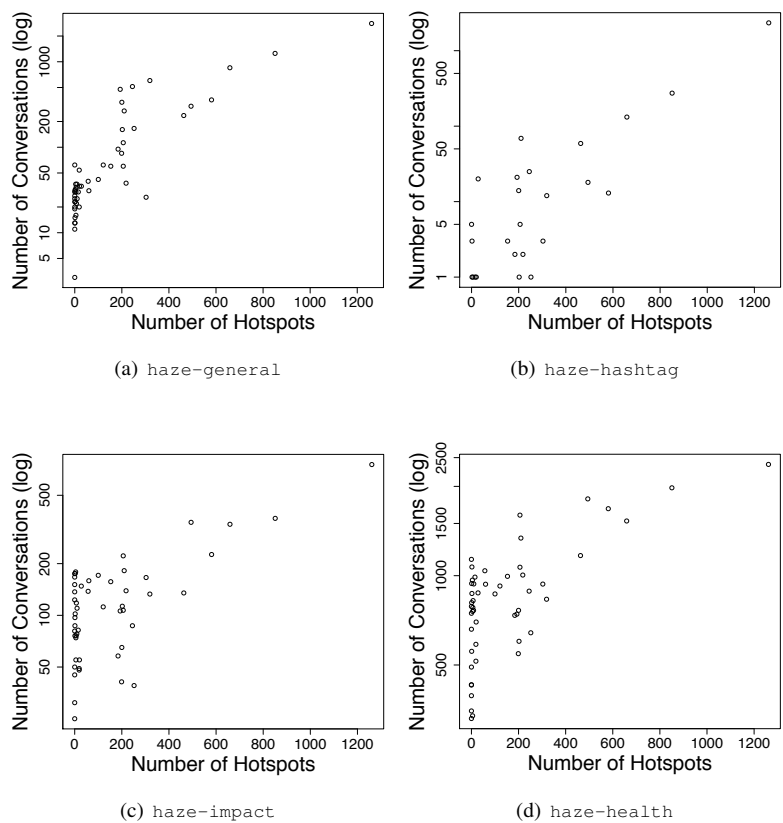

(a) `haze-general`

(b) `haze-hashtag`

(c) `haze-impact`

(d) `haze-health`

**Fig. 4: Number of detected hotspots per week vs. number of conversations (log-scaled)**

We start with an analysis of the corresponding or nearest hotspot of each tweet. The users usually discuss haze and its derived problems such as impact, rather than hotspots themselves, as hotspots exist far away and are invisible. However, those hotspots are the origin of the problem. Since people who live closer to a hotspot will be affected by haze than ones who live farther, we use a simple model that considers the distance from tweets to their nearest hotspot and analyse the corresponding or nearest hotspot of each tweet. Let us say that, given a hotspot `h`, its popularity `popularity(h)` denotes the number of tweets correspondent to `h` when the hotspot and tweets are identified and posted at the same day, respectively.

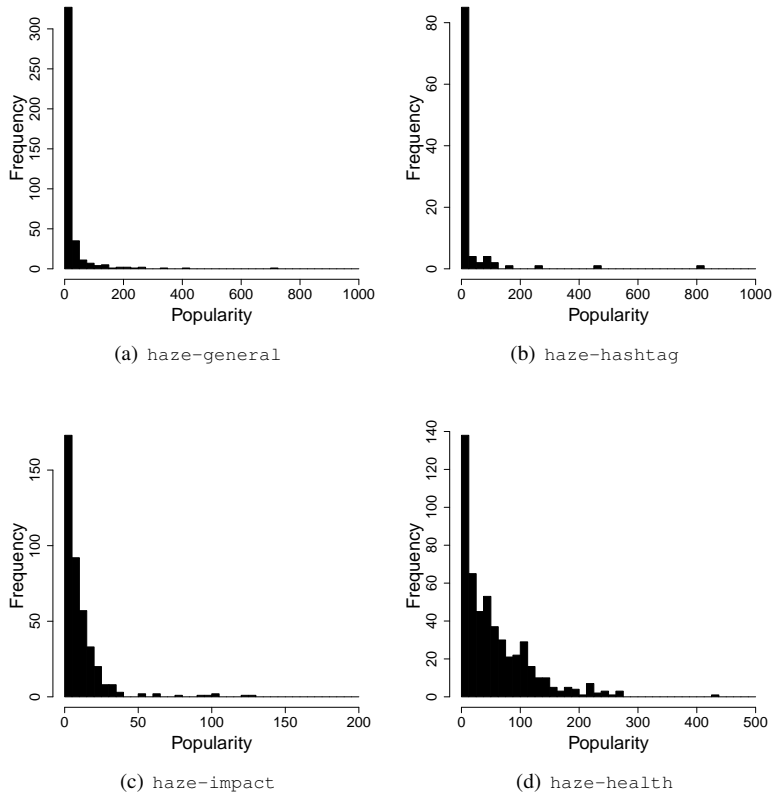

(a) `haze-general`            (b) `haze-hashtag`

(c) `haze-impact`            (d) `haze-health`

**Fig. 5: Popularity distribution of hotspots considering different taxonomies**

Figure 5 shows the frequency of `popularity(H')` where `H'` is a set of hotspots satisfying `popularity(h)` $\geq$ 1.The figure shows that (a) there are highly referenced hotspots from `haze-general` and `haze-hashtag`, such as 700-800 tweets per hotspot in some instances, while many other hotspots are referred to by only a small number of tweets from the two datasets, and (b) `haze-impact` and `haze-health` do not display as skewed distributions as the other two datasets but they still contain such characteristics.

Such an analysis is useful for a disaster management, particularly by helping prioritise hotspots based on an understanding on which hotspots are more discussed by affected people.

In the following, we ground our exploration in two sets of cascading research questions.

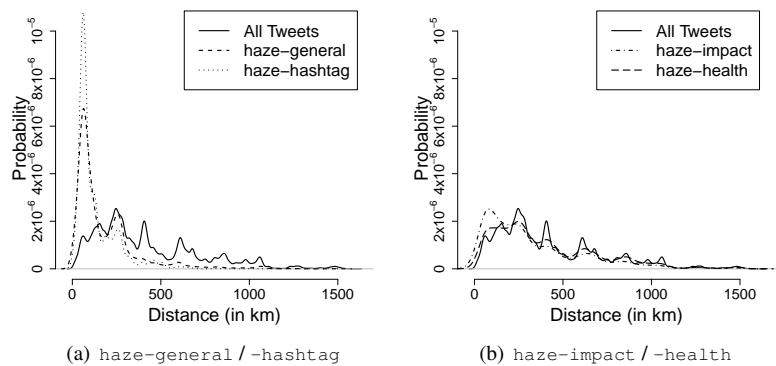

(a) `haze-general` / `-hashtag`     (b) `haze-impact` / `-health`

**Fig. 6: Probability Density Functions of the distance from conversations of given types to the nearest relevant hotspots**

**(1) What insights can be generated when we observe the locations of haze-related tweets? What is the likelihood of hotspots nearby when we observe such tweets by their known locations?**

For every tweet from the entire dataset, regardless of its haze-topic, we calculate the shortest distance between it and its corresponding (nearest) hotspot from the same day. We present its empirical probability density function of the whole distance values with straight lines in Figure 6. This is a basis from which to determine whether another distance distribution from a haze-specific topic has similar or different characteristics. Then we conduct the same process for each of the four haze datasets, represented with dotted lines in the figure.

The two distributions from `haze-general` and `haze-hashtag` are significantly different from the ones from `haze-impact` and `haze-health`, while the latter two distributions are relatively similar to the distance distribution from the entire data. Notable spikes are present in Figure 6(a) around $x$=100km. These imply that `haze-hashtag` and `haze-general` are likely posted with relatively close proximity, of up to 100km, to hotspots as local issues that attract the attention of populations living near hotspots, but `haze-impact` and `haze-health` are topics of relevance to many residents across Sumatra island. Also, a close look at the four subject-matter distributions, comparing them with the global distribution, highlights that `haze-hashtag` is the topic with the closest proximity to real world hotspots, followed by `haze-general`, `haze-impact`, and `haze-health` in a decreasing order, getting toward general topics.

**(2) What insights can be generated when we know the locations of hotspots? Could we guess what topics would be more likely to be discussed nearby?**

For this, on a given day, for every hotspot, we identify its corresponding tweet for `haze-general`, `haze-hashtag`, `haze-impact`, or `haze-health` by identifying the nearest tweet. We present the distance distribution between each hotspot and corresponding tweet along with null models in Figure 7 presented in Probability Density Functions. Note that this differs from the previous section where we investigated, given a tweet, how far it is located to its nearest hotspot.

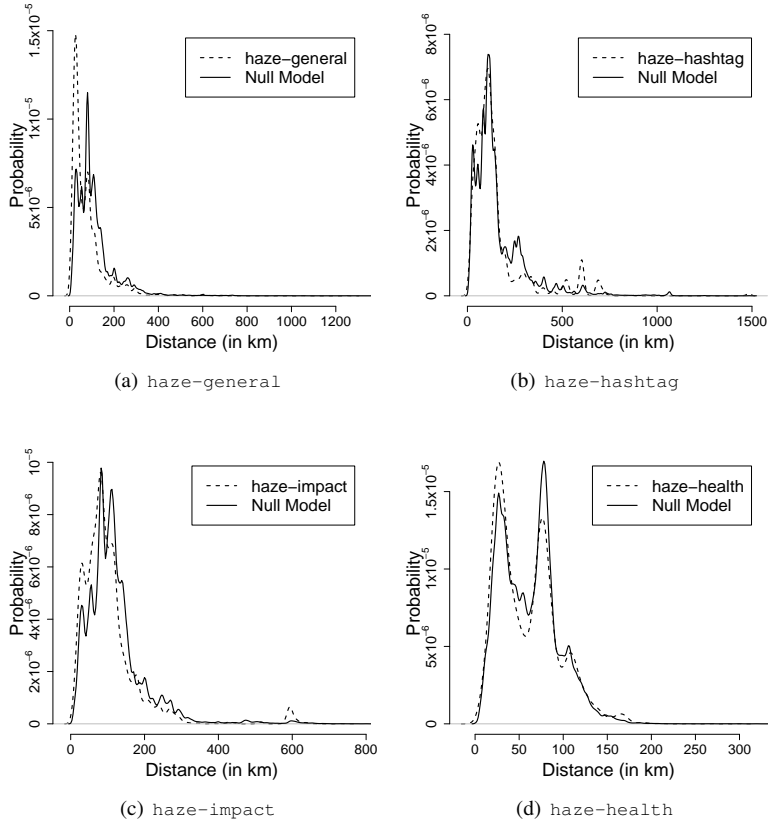

(a) `haze-general`          (b) `haze-hashtag`

(c) `haze-impact`           (d) `haze-health`

**Fig. 7: Probability Density Functions of the distance from relevant hotspots to the nearest conversations of given types**

As a null model for a haze-specific topic (for instance `haze-general`), for each date we randomly select a number of tweets, of the same magnitude as the size of the haze-specific topic compared to the total tweets posted at the same date. For instance, if $n$ tweets from given category (e. g., `haze-general`) and $m$ hotspots were detected on a given day, we choose $n$ random tweets $T$ from the whole dataset and compute distances from the hotspots to the nearest tweets from $T$. The final null model for a topic is built from multiple random processes (1,000 times). Numbers of tweets of each category is different on different days, so each haze-specific topic may have its own null model.

**Table 4: Parameters of probability distribution of the distance from hotspots to the nearest conversations of given types**

| | Average | | Median | | St.Dev. | |
|---|---|---|---|---|---|---|
| | Real Data | Null Model | Real Data | Null Model | Real Data | Null Model |
| `haze-general` | 77.1 | 109.2 | 51.5 | 87.5 | 81.5 | 88.4 |
| `haze-hashtag` | 159.3 | 161.6 | 111.8 | 118.7 | 111.8 | 144.7 |
| `haze-impact` | 104.8 | 121.4 | 85.6 | 106.9 | 88.3 | 90.9 |
| `haze-health` | 59.5 | 61.7 | 55.0 | 61.1 | 34.9 | 32.7 |

As expected from the previous two figures, the four distributions from the four different topic datasets are dissimilar. In a general sense, for all haze-specific topics (possibly except `haze-general`), the real distribution and the distribution from its null model are generally similar to one another. This implies that the creation of a hotspot is not a trigger for haze-related discussions, regardless of its geographical location. A detailed look at the distributions, however, along with a reexamination of Table 4 showing mean, median and standard deviation values, reveals a set of insights, including (a) the increased likelihood of tweets on topics including `haze-general` and `haze-impact` closer to a hotspot compared to their null models and (b) the greater density of tweet locations on `haze-hashtag` of a given hotspot, compared to a null model.

It is worth noting that the two research questions and their results we discuss in Section 4.2 are complement each other and useful in different aspects in disaster management. First, we showed that `haze-general` or `haze-hashtag` conversations can be indication of a nearby hotspot. Health and haze impact issues are discussed all over Sumatra Island, not necessary near hotspots. In the second question, we tried to identify relevant tweets near the hotspots. There are only few "popular" hotspots with many tweets near them. The reason may be that many hotspots are detected in rural areas.

## 5 Change of Mobility Analysis

Peat fires and haze incidents have an impact on people's behaviour, especially their mobility patterns. Some residents move to haze-free areas, while others stay at home but limit their outdoor activities. In rare cases, the government advises an evacuation due to severe haze and the need to protect affected and vulnerable populations. In this section, we investigate mobility patterns, which is of practical importance as often this information but is not available in a (near) real-time fashion which would be preferable for disaster management. We tackle this by looking at the spatial-temporal information available from social media and investigate how patterns of mobility artuculated by social media can be quantified. For this, we analyse all the tweets with GPS information posted by users who spend most of their time in Riau Province, one of the most haze-affected regions, as explained by Figure 2(a).

## 5.1 Mobility in General

For this part of the study we first classify all weeks in 2014 into three categories based on the weekly number of hotspots in Sumatra as shown in Table 5, with the exception of weeks which see large population movements such as over new year, school holidays, and religious celebrations. Then we add a special week, $W^E$, during which an evacuation was advised by a local government (the Week of March 13, 2014). This classification is easily extended to the level of users, $W_u^{NH}$, $W_u^H$, $W_u^{HS}$, and $W_u^E$, by only examining a week when the number of tweets during the week posted by User $u >$ a threshold $\tau \in \mathbb{N}$. We use $\tau = 4$, but when using different threshold values, we have similar results.

**Table 5: Four types of weeks by the number of weekly hotspots (`#w.h.`) in Sumatra**

|          | Description | Condition | Week |
|----------|-------------|-----------|------|
| $W^{NH}$ | `no-haze`      | `#w.h.¡100`       | 2, 18–24, 33–36, 43, 46–51 |
| $W^H$    | `haze`         | `100¡#w.h.¡400`   | 5, 6, 12, 37–42, 44, 45 |
| $W^{SH}$ | `severe-haze`  | `400¡#w.h.`       | 7–10, 11, 13 |
| $W^E$    | `evacuation`   | `—`               | 11 |

Now we define four notations as presented in Table 6 with specific reference to all users $C^w$, $S^w$, $D^{(w1,w2)}$, and $RS^{(w1,w2)}$.

**Table 6: Four notations to analyse mobility patterns**

| | |
|---|---|
| $C_u^w$ | **CENTROID** (average latitude and average longitude) of all GPS-stamped tweets posted by User $u$ in the Week $w$ |
| $D_u^{(w1,w2)}$ | Euclidean **DISTANCE** between two centroids, $C_u^{w1}$ and $C_u^{w2}$ |
| $S_u^w$ | **SPREAD**, the average distance between $C_u^w$ and tweets posted by $u$ in Week $w$ |
| $RS_u^{(w1,w2)}$ | **RELATIVE SPREAD** between two weeks calculated by $S_u^{w2} / S_u^{w1}$ |

**(1) Mobility Patterns by Situation (Analysing DISTANCE)**

During a slow onset disaster situation, an affected person can change her or his mobility pattern. The individual has two options to (significantly) increase her or his movements due to an evacuation or to (significantly) decrease her or his mobility by staying at home. In order to quantify changes in mobility, we calculate the distributions of DISTANCE $D_u^{(w1,w2)}$ when $w1 \in W_u^{NH}$ and $w2 \in \{W^{NH}, W^H, W^{SH}, W^E\}$ as shown in Figure 8.

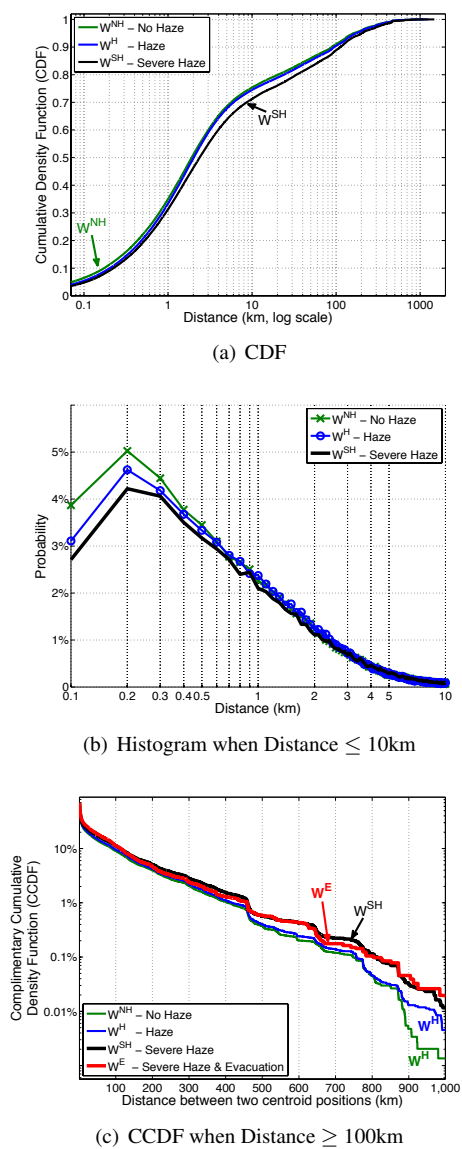

(a) CDF

(b) Histogram when Distance $\leq$ 10km

(c) CCDF when Distance $\geq$ 100km

**Fig. 8: Distributions of the numbers of users classified by the distance between two centroid positions per situation**

Figure 8(a) shows the overall distriburion of $D_u^{(w1,w2)}$ using Cumulative Density Functions (CDFs) for three cases except $w2 \in W^E$. Overall people tend to move further during a severe haze period, while mobility patterns during periods of (light) haze and no haze are very similar, as implied by the observation that the two distri-butions, $W^{NH}$ and $W^{H}$, are similar to each other while the distribution for $W^{SH}$ is stretched toward the x-axis in the positive direction.

Considering smaller order mobility of up to 10km, shown in Figure 8(b), we see that during the haze free weeks ($w2 \in W^{NH}$) residents make more short-distance movements than during the weeks with haze, $w2 \in \{W^{H}, W^{SH}\}$.

From the tails of the distance distributions in Figure 8(c), an investigation of long distance movement patterns shows a tendency that some people significantly increase their mobility by hundreds of kilometers during evacuation periods and severe haze $w2 \in \{W^{SH}, W^{E}\}$, while there is no significant difference between the (light) haze and no haze periods, $w2 \in \{W^{NH}, W^{H}\}$.

**(2) Mobility Reduction by Situation (Analysing DISTANCE and RELATIVE SPREAD)**

Now we focus on the reduction of mobility. Examining a week from among haze-free weeks (say $w1 \in W^{NH}$), we interpret that a user (say $u$) reduces her or his mobility in another week (say $w2$) when $u$'s RELATIVE SPREAD is smaller than 1 or $RS_u^{(w1,w2)} < 1$. In this article, we specifically say that $u$ does reduce her or his mobility if $RS^{(w1,w2)} < 1/3$ which means she or he reduced her or his SPREAD more than three times [4].

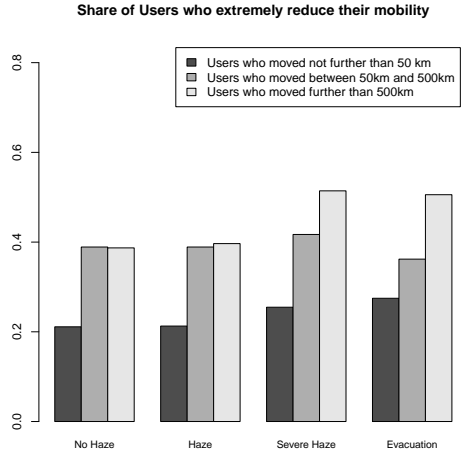

**Fig. 9: Proportions of users who reduce mobility compared to `no-haze` weeks with respect to haze situation and the distance people moved away**

Additionally, the information on where people currently reside is important to understand movement patterns. For instance, during a period with severe haze, people who stay in an affected area may display a different pattern of mobility reduction when compared to people who have already evacuated from the area.

---

[4] We tried different threshold values (e. g., $1/5$), but the results were similar.

Figure 9 summarises people's mobility behavior changes given two different weeks in different circumstances, classified by DISTANCE $D^{(w1,w2)}$ and our haze/non-haze scenarios. Of note, the baseline weeks remain as the haze-free week scenario (*i.e.*, w1 $\in W^{NH}$). In other words, we measure, for a given user, to what extent she reduces her mobility in a haze-free week, a haze week and a severe haze week, compared to a baseline haze-free week. For instance, the very first bar shows that about 20% of users, out of the users who move less than 50km in two haze-free weeks, reduce their mobility, although their mobility patterns are studied in two haze-free weeks. It means that overall about 20% of users exhibit their mobility patterns within a wide range of variation, even when they still remain in a city (e.g. within a 50km radius).

This figure shows a couple of interesting points. First, considering people who do not move far from their previous location, the probability that they noticeably decrease their mobility (i.e. by a factor of three) is higher during severe haze weeks and evacuation periods (about 25% and 27%, respectively) than during haze free and moderate haze weeks (about 21%). A similar result will be discussed in Section 5.2-(2). Second, people that move far away, similar to the above, also reduce their mobility in the haze and evacuation periods (about 51.4% and 50.6%, respectively) compared to the no haze or moderate haze periods (about 38-39%) but the proportion is much larger. The above may imply that although some people move far away (*e.g.,* more than 500km), expecting fresh air, their movement patterns in their new location are decreased. One possible explanation is that moving to an new area naturally reduces mobility due to the unfamiliarity of the surroundings.

## 5.2 Mobility Patterns with Ground-truth Situational Data in Riau

The importance of this section is based on a close interpretation of mobility-related information revealed from social media and real-world situations including air quality, local context such as infrastructure-based connectivity, and people's custom behaviors, to qualitatively assess the potential of social media as a complementary data source. With this purpose, we analyse mobility patterns, taking into account the administrative boundaries in Riau province, of a set of users (838 users) whose home locations are identified as Pekanbaru, the capital city of Riau province and the city with the largest populations in Riau province[5] , as well as their tweets (109,096 GPS-stamped tweets).

Riau consists of 12 cities and regencies, each of which is identified by a 4-digit post code; a regency or a city consists of sub-districts. For a detailed study, in this section, we limit the timespan to include February 26 and March 23, which is exactly when the information on air quality, officially published by the Indonesian National Board for Disaster Management [6], is available[6]. Based on the air quality informa-

---

[5]Indonesian Central Bureau of Statistics

 http://riau.bps.go.id/linkTabelStatis/view/id/210

[6]http://www.menlh.go.id/DATA/ispu_riau.PDF

**Table 7: Air quality of different regencies in Pekanbaru between February 26 and March 23**

| Postal Code | February | | | March | | | | | | | | | | | | | | | | | | | | | | |
|---|---|---|---|---|---|---|---|---|---|---|---|---|---|---|---|---|---|---|---|---|---|---|---|---|---|---|
| | 26 | 27 | 28 | 1 | 2 | 3 | 4 | 5 | 6 | 7 | 8 | 9 | 10 | 11 | 12 | 13 | 14 | 15 | 16 | 17 | 18 | 19 | 20 | 21 | 22 | 23 |
| 1401 | – | – | – | – | – | – | – | – | – | – | – | – | – | – | – | – | – | – | – | – | – | – | – | – | – | – |
| 1402 | – | – | – | – | – | – | – | – | – | – | – | – | – | – | – | – | – | – | – | – | – | – | – | – | – | – |
| 1403 | – | – | – | – | – | – | – | – | – | – | – | – | – | – | – | – | – | – | – | – | – | – | – | – | – | – |
| 1404 | – | – | – | – | – | – | – | – | – | – | – | – | – | – | – | – | – | – | – | – | – | – | – | – | – | – |
| 1405 | – | B | – | B | – | B | B | B | B | B | B | B | B | B | B | B | B | R | Y | BL | BL | Y | Y | Y | BL | BL |
| 1406 | – | – | – | – | – | – | – | – | – | – | – | R | Y | Y | B | B | Y | G | G | – | G | BL | G | G | BL | |
| 1407 | – | – | – | – | – | – | – | – | – | – | – | – | – | – | B | Y | | | | | | | | | | |
| 1408 | B | B | R | B | B | B | R | Y | R | Y | B | B | B | R | B | B | B | B | BL | BL | BL | G | BL | BL | BL | |
| 1409 | – | B | B | B | B | B | B | B | B | B | B | B | B | B | BL | BL | BL | BL | Y | Y | – | Y | BL | | | |
| 1410 | – | – | – | – | – | – | – | – | – | – | – | – | – | – | – | – | – | – | – | – | – | – | – | – | – | – |
| 1471 | Y | Y | Y | Y | Y | Y | Y | Y | Y | Y | R | B | R | R | B | B | B | Y | BL | BL | BL | G | BL | G | G | BL |
| 1473 | B | Y | B | Y | R | B | B | B | R | B | B | B | B | R | Y | B | B | B | G | BL | BL | BL | Y | BL | BL | G |

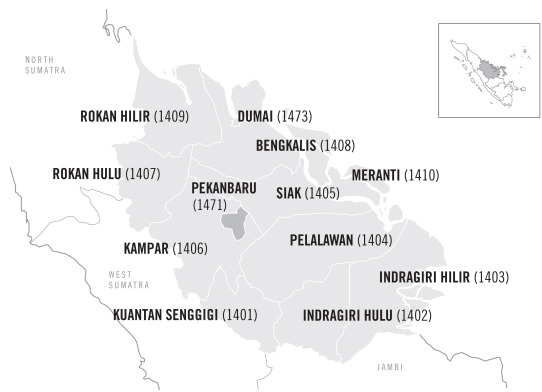

**Fig. 10: Map of Riau province with postal codes**

tion, each day in each regency (identified by 4-digit post code) can be associated with one of the classes of air quality: {'**G**'reen, '**BL**'ue, '**Y**'ellow, '**R**'ed, '**B**'lack}, where '**G**'reen means the best air quality, and '**B**'lack signifies the worst air quality. Table 7 shows daily air quality in 12 different regencies between February 26 and March 23. Figure 10 shows the locations of those regencies.

**(1) Diversity of Users' Mobility**

First, we try to estimate the overall mobility of citizens of Pekanbaru, the capital of Riau Province. Table 8 consists of three parts. The header contains the information about air quality in Pekanbaru. The second part, Table 8[1], contains daily numbers of regencies that citizens of Pekanbaru visit. The third part, Table 8[2], shows how users move inside of different regencies – particularly, how many sud-districts were visited in each regency or city. Figure 10 shows the locations of the regencies in Riau Province.

We use the number of cities and regencies visited by them as a measure of their movement diversity. Table 8[1] shows how many different cities and sub-districts in-side and outside of Riau were visited by those users. The diversity of visited regencies is consistent most of the time, except during the period following an evacuation recommendation by the local government. In that period users visit many more areas, possibly trying to escape the haze. This type of information, namely whether citizens listen to and act on evacuation advice, is not currently captured by the government in a real-time manner.

**Table 8: Distribution of regencies and sub-districts, visited by Pekanbaru residents between Feb 26 and Mar 23 (best viewed in color):**

[1] **(a-b) Number of regencies and cities visited by Pekanbaru citizens outside (a) and inside (b) Riau**

[2] **Number of sub-districts visited by Pekanbaru residents in each regency**

| Date | | Feb 26 | 27 | 28 | Mar 1 | 2 | 3 | 4 | 5 | 6 | 7 | 8 | 9 | 10 | 11 | 12 | 13 | 14 | 15 | 16 | 17 | 18 | 19 | 20 | 21 | 22 | 23 |
|---|---|---|---|---|---|---|---|---|---|---|---|---|---|---|---|---|---|---|---|---|---|---|---|---|---|---|---|
| Air Quality (Pekanbaru) | | Y | Y | Y | Y | Y | Y | Y | Y | Y | Y | R | B | R | B | B | B | B | B | Y | BL | BL | G | BL | G | BL | BL |
| [1] | (a+b) | 6 | 5 | 7 | 8 | 5 | 3 | 2 | 3 | 2 | 3 | 5 | 4 | 4 | 6 | 4 | **9** | **10** | **10** | **12** | 6 | 2 | 3 | 3 | 5 | 4 | 4 |
| | (a) | 2 | 2 | 1 | 3 | 1 | 1 | 0 | 1 | 1 | 1 | 1 | 2 | 0 | 0 | 0 | **2** | **4** | **1** | **5** | 2 | 0 | 0 | 0 | 1 | 0 | 1 |
| | (b) | 4 | 3 | 6 | 5 | 4 | 2 | 2 | 2 | 1 | 2 | 4 | 2 | 4 | 6 | 4 | **7** | **6** | **9** | **7** | 4 | 2 | 3 | 3 | 4 | 4 | 3 |
| [2] | 1401 | 1 | | | | | | | | | | | | | | | | 2 | | 1 | | | | | | | |
| | 1402 | 1 | | 1 | | | | | | | | | | 1 | 1 | 1 | 1 | 2 | 2 | 1 | | | | | | | |
| | 1403 | | | | | | | | | | | | | | | 1 | 1 | 2 | | | | | | | | | |
| | 1404 | | | 1 | 1 | | | | | | | | | | | | | 1 | | 1 | | | | | | | |
| | 1405 | | | | | | 1 | 1 | | | | | | | | | 2 | 2 | 1 | 2 | 1 | 1 | 1 | 1 | | | |
| | 1406 | 1 | 2 | 1 | 2 | | | | 1 | 1 | | 1 | 1 | 2 | | 1 | 2 | 5 | | 1 | 1 | | 2 | | | 1 | 1 |
| | 1407 | | | | 1 | 1 | 1 | | | | | | | | 1 | 1 | 2 | 4 | | | | | | | | | |
| | 1408 | | 1 | 1 | 1 | 1 | | | | | | 1 | | 1 | 1 | 2 | 1 | 1 | 1 | 1 | 1 | | | | | | |
| | 1409 | | | | | | | | | | | | | | | | | | | 1 | | | | | | | |
| | 1410 | | | | | | | | | | | | | | | | | | | | | | | | | | |
| | 1471 | 3 | 4 | 5 | 5 | 2 | 4 | 4 | 2 | 2 | 2 | **4** | **3** | **2** | **2** | **2** | **3** | **4** | **4** | 3 | 3 | 2 | 4 | 4 | 3 | 3 | 3 |
| | 1408 | | | | | | | | | | | | | | | 1 | | | | 1 | | | | | | | |

**(2) Mobility Behaviors of Pekanbaru Citizens**

Next, we demonstrate how social media enables us to recognise the behavioral patterns of users. We use air quality data and other information to try to interpret the identified patterns. This enables us find different insights in social media data. Table 8[2] shows the daily number of sub-districts that Pekanbaru residents visit, based on tweet locational information. For instance, no one who live in Pekanbaru tweets (or visits) in regency `1410` during the period of the study.

- *'Moving to Areas with Cleaner Air'*: During the severe haze situation in Pekanbaru, from March 8 till March 15, users tend to visit regencies in the South (`1401`, `1402`, `1403`) and the West (`1406`, `1407`). Southern and western regencies had better air quality than other regencies in Riau, according to information from the media. We also observe more activity in the regency `1408` although the air quality in `1408` is similarly bad to that of Pekanbaru. An investigation confirms that they were present only on the main road of Sumatra island in `1408` which connects to western regencies.

- *'Avoiding Areas with Bad Air Quality'*: The citizens of Pekanbaru avoided visiting the regency `1405` until the air quality improved. Another example of how people avoid areas affected by haze are the regencies `1409` and `1473`. These regencies were affected by haze during the recommended evacuation period, and users also avoid these areas during that period of time.

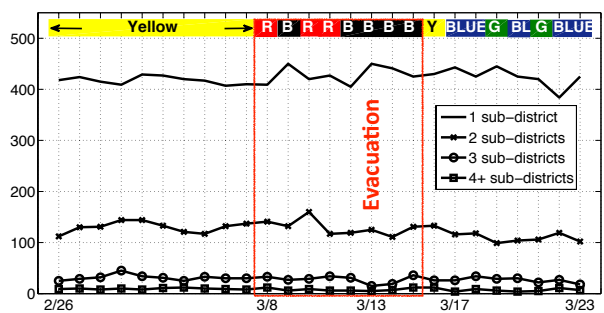

**Fig. 11: Number of users who visit one, two, three, or over four sub-districts per day in Pekanbaru**

- *'Limited Mobility in Pekanbaru'*: According to the Table 8[2], regency `1471` has not shown an outstanding dynamical change of the number of sub-districts, while a tendency is observed that people limited mobility before and during the severe haze period. For a closer look, we count the number of users who posted in one, two, three, and more than four sub-districts in regency `1471`. In Figure 11, we find a tendency that the number of users who visit one sub-district per day increases and the number of users who visit two more sub-districts slightly decreases during the severe haze period, and especially after the evacuation order.

### 5.3 Additional Information related to Mobility Patterns

In this section, we study mobility patterns by analyzing tweets posted by the users used in previous section, from two different angles, what (possibly) people say and what (possibly) people do. Concerning what people say, we collect and analyze the dynamics of tweets over time mentioning two words, 'home' (*rumah*) and 'evacuation' (*evakuasi*, *ngungsi*). Regarding what people do, we identify two lists of tweets posted (i) by Foursquare[7] and (ii) using web browsers, instead of via a mobile application, which is part of the meta-information of a tweet, and present its dynamics in Figure 12. Again similar to the results discussed in Section 5, this does not aim to quantify the exact number of people who stay in or evacuate from affected areas but aims to understand whether this information could show any tendency or behaviour change.

First, two thin lines, depicted in red, show what people say. There are peaks around March 13 when the government advised residents to evacuate. The two thicker lines, depicted in black, signify behaviour. The number of tweets posted by regular web browsers clearly increases around March 13, while the number of Foursquare check-in around the same period of time slightly decreases. We would be able to reference this information to better understand haze-affected residents' behaviors and to complement existing and new types of information for disaster response, but this should be further validated and evaluated which we leave for future research.

---

[7] `http://foursquare.com`

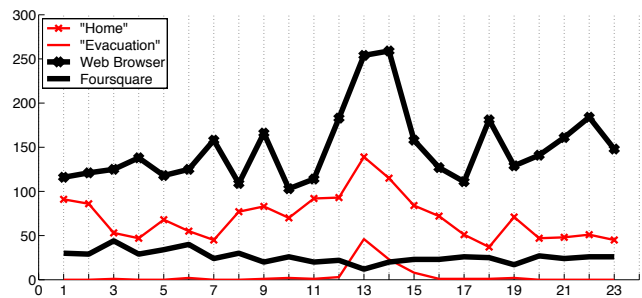

**Fig. 12: Numbers of tweets that (a) mention 'home' or 'evacuation' or (b) are posted using web browsers or from FOURSQUARE**

## 6 Discussion - Applications, Challenges, and Collaborations

We have mined and analysed social media to test its potential to address a real-world problem, namely the need to provide (near) real-time situational information on peatland fire and haze disasters in Indonesia, as complementary information. Meanwhile we validated our approach and methods with secondary information, such as quantitative data, namely hotspot and air quality, contextual information, including typical evacuation behaviors led by geographical characteristics, and qualitative discussions, namely consultations with officials involved in humanitarian activities in Indonesia, despite limited access to disaster-related data due to the nature of haze disasters. In this chapter, we discuss applications and challenges, by examining other information needs in local and global contexts, briefly explaining a platform we are building for policy-makers based on the results of this analysis, and reviewing research challenges for the attention of the research community.

### (1) Information Needs

Humanitarian action can be characterised by three stages, `before`, `during` and `after` an event, where the `before` stage involves disaster preparedness and the development of early warning systems, the `during` stage deals with humanitarian needs during a particular disaster, and the `after` stage is related to disaster recovery [8]. Even though data and information exist and are commonly used throughout all states[8] such as a population distribution as baseline information, different sets of data and information are required at each stage. Recently social media has been investigated in both an active mode as a tool for collective knowledge, for instance to map affected areas, assess damage and disseminate information quickly, and in a passive mode such as mining tweets in a humanitarian theatre, for instance to improve an early warning system and better understand supply and demand, as we mentioned in Section 2.

---

[8]For instance, IASC Guidelines Common Operational Datasets (CODs) in Disaster Preparedness and Response (Visit - `www.humanitarianresponse.info`)

Moreover, last year the United Nations adopted the Sendai Framework[9], a non-binding agreement for Disaster Risk Reduction as the successor of the Hyogo Framework, to better prepare for different natural disasters by providing global guidelines. This framework encourages national disaster management authorities to prioritise national policy agendas for understanding disaster risks and to strengthen disaster risk governance. But some countries, including some developing countries, are experiencing difficulties in implementing aspects of the guidelines, due in part to missing information, some of which, our and other research suggests to extract or mine from new digital data sources including social media.

**(2) A system in progress**

Throughout this article we see that social media can provide useful information, such as a better understand of residents' behavioral changes over time. Section 4 addresses opportunities for `before`, `during` and `after` haze disasters, and Section 5 shows opportunities for `during` and `after` haze disasters. We consulted with officials who have been involved in haze disasters who confirmed that this type of information will be useful for disaster management, but its utility requires further clarification and has to be improved upon taking into account inputs from local and national governments as well as UN agencies. Pulse Lab Jakarta is currently building a system to collect, analyze and visualize relevant data and information including from social media, which was discussed in this article, not only for providing complementary and supplementary information about haze and peatland fires, but also for assisting in the coordination process between different public bodies, who maintain different datasets relevant to disaster response.

**(3) A common research agenda**

In this article, we presented the results of analysis connected to a specific disaster, namely haze on the scale witnessed in Indonesia, but from limited angles, for example conversations and mobility patterns, while analyzing GPS-stamped social media data. When we broaden the scope to other types of disasters and their associated research angles and new types of information, many more practical and theoretical research questions arise, but few have been addressed to date by the research community. Social media is a data source that many researchers have recently used for quantitative and qualitative studies of human beings and societies. Clearly, using social media in a passive mode is a task with many challenges, such as whether mined information is correct or not, but we see many opportunities to help (potentially) affected people and disaster management authorities across all three stages of humanitarian action. This article aims to share these findings with the research community but more importantly also aims to refresh this area among researchers and request a collaboration between the research community and the humanitarian sector. UN Global Pulse wishes to foster a stronger link between domain specific challenges and research methods from other disciplines.

---

[9] http://www.unisdr.org/we/coordinate/sendai-framework

## 7 Summary and Future Work

Haze and peatland fires remain a near annual disaster in Indonesia and South-East Asia. As the phenomenon affects millions of citizens in Indonesia and beyond, disaster management needs significant improvement. We propose to use social media (Twitter) for haze disaster management by extracting complementary and supplementary information. We showed how users of Twitter react to haze emergencies and found correlations between the public discourse on Twitter and peat fire hotspots. We also showed how it is possible to understand the changes in users' travel patterns during haze periods using social media. We demonstrated that, despite some limitations, social media data can be used to inform disaster management at different stages of an emergency. We aim to encourage additional work and research in this field. Concerning our own research agenda, we hope to extend our data sources and to offer data-driven solutions for the management of haze disasters.

### Acknowledgement

We thank Johan Kieft from UN environment who provided insight about haze disasters in Indonesia and George Hodge from Pulse Lab Jakarta for his assistance. Furthermore, we acknowledge the use of FIRMS data and imagery from the Land, Atmosphere Near real-time Capability for EOS (LANCE) system operated by the NASA/GSFC/Earth Science Data and Information System (ESDIS) with funding provided by NASA/HQ.

## Appendix

**Table 9: English translation of the filtering rules for the identification of corresponding tweets, see Table 2**

| | |
|---|---|
| **haze-general** | **Conversations about forest and peat fires and haze, detected primarily by the keywords — 9,707 tweets** (e.g., *"When the haze problem will be solved?"*):<br><br>`( (disaster‖storm‖pollution‖severe‖thick‖fog) && (haze)`[*]`)`<br>`OR ( (danger‖alert‖emergency‖strong‖heavy) && (haze)`[*]`)`<br>`OR ( haze‖fire spot‖fire source‖hotspot‖air pollution‖haze`[†]`)`<br>`OR ( forest fire`[†]` ) OR ( (damage‖logging`[‖]`‖open‖fire`<br>`‖deforestation‖logging`[‖]`) && (forest‖field‖land‖peat) )` |
| **haze-hashtag** | **Conversations which contain one of identified hashtags — 3,024 tweets** (e.g., *"Let's participate in #melawanasap movement."*):<br><br>`( #saveriau‖#prayforriau‖#melawanasap‖#prayforasap‖#hentikanasap )` |
| **haze-impact** | **Conversations about happenings in a negative way due to haze, such as flight delay or school closing — 6,994 tweets** (e.g., *"Day #3 off because of haze."*):<br><br>`( (close‖cancel‖cacelled‖delay‖delay`[†]`‖cancel`[†]`‖closed)`<br>`&& (flight‖airport) ) OR ( visibility )`<br>`OR ( school‖college‖university) && (closed‖close‖leave‖left) )`<br>`OR ( (economy‖impact‖effect‖loss‖down‖risk) && (haze)`[*]`)` |
| **haze-health** | **Conversations with keywords indicating haze-related or derivable diseases — 46,241 tweets** (e.g., *Welcome to Pekanbaru; do not forget to wear mask!"*):<br><br>`( (infection‖stertorous) && (respiratory‖breath‖breathing) )`<br>`OR ( (irritation‖inflammation) && (eye‖skin‖throat‖nose‖lungs`[‡]`) )`<br>`OR ( cough‖headache‖sick‖ari`[§]`‖mask‖asthma‖asthma`[†]`‖lungs`[‡]`)`<br>`OR ( (haze)`[*]`&& (health‖healthy‖breath‖pregnant‖child`<br>`‖elderly) ) OR ( eye && (sore‖irritate‖acute)`[¶]`)` |

---

[*]"Kabut asap" is translated as "haze". Both words "kabut" and "asap" are also often used to describe haze. However, if used not together, both words also have further meanings.

[†]These keywords were originally used in English.

[‖]Both words "pembalakan" and "penebangan" are synonymous and mean "logging".

[‡]"Paru-paru" is translated as "lungs". Sometimes, the word "paru" is translated as "lungs" as well. The word "paru" is ambiguous unless it is put in context with further words, such as "infection".

[§]ARI stands for Acute Respiratory Infection. We used the abbreviation ISPA (Infeksi Saluran Pernafasan Akut) which is commonly used by Indonesian Twitter users.

[¶]All three terms "pedih", "perih", "sakit" describe the eye sore.